%% file: conference_101719.tex
\documentclass[conference]{IEEEtran}
\IEEEoverridecommandlockouts

\usepackage{graphicx}
\usepackage{caption}

\usepackage{cite}
\usepackage{amsmath,amssymb,amsfonts}
\usepackage{algorithmic}
\usepackage{graphicx}
\usepackage{textcomp}
\usepackage{xcolor}
\def\BibTeX{{\rm B\kern-.05em{\sc i\kern-.025em b}\kern-.08em
    T\kern-.1667em\lower.7ex\hbox{E}\kern-.125emX}}
\begin{document}

\title{Logical Scalability in Big Data Architectures: Impact of Data Modeling on the Database Layer
}

\author{
\IEEEauthorblockN{Rosario Napoli}
\IEEEauthorblockA{\textit{University of Messina}\\
Messina, Italy \\
0009-0006-2760-9889} 
\and
\IEEEauthorblockN{Mark Adrian Gambito}
\IEEEauthorblockA{\textit{University of Messina}\\
Messina, Italy \\
0009-0008-3706-4493} 
\and
\IEEEauthorblockN{Antonio Celesti}
\IEEEauthorblockA{\textit{University of Messina}\\
Messina, Italy \\
0000-0001-9003-6194} 
\and
\IEEEauthorblockN{Maria Fazio}
\IEEEauthorblockA{\textit{University of Messina}\\
Messina, Italy \\
0000-0003-3574-1848} 
}

\maketitle

\input{tex/abstract}

\input{tex/problem_identification}

\input{tex/related_works}

\input{tex/motivation}

\input{tex/material_and_methods}

\input{tex/results}

\input{tex/discussion}

\input{tex/reproducibility}

\section*{Acknowledgments}
Rosario Napoli is a PhD student enrolled in the National PhD in Artificial Intelligence, XL cycle, course on Health and life sciences. Mark Adrian Gambito is a PhD student enrolled in the National PhD in Artificial Intelligence for Society, XXXIX Cycle. 

\bibliographystyle{unsrt}
\bibliography{bib}

\end{document}

%% file: tex/abstract.tex
\begin{abstract} 
In modern Big Data architectures, scalability is often addressed through infrastructural expansion, involving vertical or horizontal scaling to manage growing data volumes. However, this hardware-centric perspective overlooks the logical scalability of the storage layer.
A central challenge lies in how data modeling choices affect the storage layer of big data architectures. Specifically, when the data model is misaligned with the underlying query patterns, performance bottlenecks emerge and persist regardless of the available computational resources.

This paper investigates the concept of logical scalability by conducting a comparative performance analysis of Document, Graph and Multi-Model database paradigms under varying workloads, queries and dataset scales. Our results demonstrate that performance degradation caused by model misalignment cannot be mitigated by additional computational resources alone. Increasing the number of machines while retaining a structurally inadequate data model yields performance equivalent to or worse than a single-machine configuration, as distributed coordination overhead compounds the inefficiencies of the underlying logical model. We demonstrate that efficiency is maximized when the logical structure of the database aligns with the inherent nature of the data and the intended analysis scope, a condition that proves to be more decisive than any infrastructural scaling strategy.
\end{abstract}

\begin{IEEEkeywords} Logical Scalability, Big Data architectures, Big Data Storage Layer, Data Modeling, NoSQL Database Paradigms.
\end{IEEEkeywords}

%% file: tex/problem_identification.tex
\section{Problem identification}
For over three decades, Relational Database Management Systems (RDBMSs) have represented the standard for data management and analytics \cite{b2}.  The relational paradigm relies on a one-size-fits-all approach to enforce strict ACID compliance (Atomicity, Consistency, Isolation, and Durability) with rigid schemas \cite{b4}. Despite its wide usage, the exponential growth in data volume, velocity and variety has progressively exposed the limitations of such an approach,  driving a paradigm shift towards Big Data, which prioritizes scalability and flexibility over traditional assumptions \cite{b3}.
Specifically, Big data systems are structured into a layered architecture spanning: physical infrastructure, data storage, organization, warehousing, analytics and application layers \cite{b1}, organizing the data lifecycle into distinct functional stages.

Among all these components, the storage layer plays a central role in enabling real-time data persistence and retrieval, by using data management techniques that are designed to meet the throughput and latency requirements of different data streams \cite{b25}.
Within this heterogeneous ecosystem, databases with different logical models, also known as NoSQL DBMS (Not Only SQL),  have emerged to address different data modeling and scalability requirements \cite{b32}. Unlike relational models, however, this domain lacks a single established standard, presenting a much more complex scenario. Generally, NoSQL DBMS are classified into several categories, specifically: column-family, document, graph and key-value stores \cite{b5}. Each category is designed to optimize specific data profiles, ranging from massive, low-complexity datasets to smaller, highly interconnected structures. Managing this variety of logical models creates challenges for Big Data analytics workflows, as real-world applications rarely align neatly with such distinct categories \cite{b7}. Common approaches to scalable computing attempt to mitigate bottlenecks through infrastructural expansion, scaling vertically or horizontally for growing workloads \cite{b8}.

However, this perspective overlooks the logical scalability of the data model. Regardless of hardware capacity, selecting an ill-suited paradigm inherently constrains the system's performance. Specifically, when the schema is misaligned with query patterns, additional computational power cannot effectively mitigate the resulting latency \cite{b9}.
This limitation becomes particularly critical in hybrid scenarios where workloads demand both the agile handling of flexible data structures and the analysis of complex relationships. While recent multi-model DBMSs support multiple paradigms within a single engine, they do not eliminate the need to align data with the most efficient logical representation \cite{b26}.
Therefore, achieving high performance depends not only on increasing computational power, but also on selecting a logical model that minimizes the mismatch between the data structure and the retrieval requirements.

To address this challenge, this paper investigates the boundaries of logical scalability, taking into account the capabilities of modern NoSQL DBMS. Specifically, we conduct a comparative analysis of Document, Graph and recent Multi-Model paradigms, selected as they represent the most widely adopted and complex models in the current landscape. We focus on quantifying the performance impact of model misalignment to provide a data-driven study for Big Data architectural decision-making.
The study utilizes a dataset comprising over 1 million nodes and 1.5 million relationships, partitioned into incremental subsets to evaluate performance variations in the storage layer across different scales and configurations, particularly:
\begin{itemize}
\item We analyze how performance varies with query complexity, demonstrating the impact of the logical model on the system's ability to scale.
\item We provide empirical evidence that indicates that efficiency is maximized when the dataset aligns with the logical structure of the database.
\end{itemize}

These findings contribute to a framework for defining logical scalability, proving that the alignment between the data model and query patterns is a prerequisite for optimizing Big Data Systems' infrastructure.

%% file: tex/related_works.tex
\section{State of the Art} \label{sec:related_work}

The evolution of Storage Layers for Big Data has shifted from rigid relational schemas towards flexible NoSQL architectures, driven by the need to handle the Volume, Variety, and Velocity of modern information flows \cite{b10}. Early scalable data management approaches relied heavily on horizontal partitioning (sharding) within Key-Value or Column-family stores to maximize write throughput \cite{b11}. These techniques were subsequently refined by the emergence of Document-oriented stores, such as Document Oriented DBMS, which prioritize high availability and schema flexibility. This enables the agile management of semi-structured data without complex pre-processing \cite{b12}. Meanwhile, the growing prevalence of highly interconnected data has led to the adoption of graph databases, which treat relationships as first-class entities to optimize traversal performance \cite{b30}.

However, most existing comparative studies focus predominantly on the infrastructural dimension of scalability \cite{b14}. Specifically, standard benchmarks such as YCSB (Yahoo! Cloud Serving Benchmark) evaluate performance based on raw throughput and latency under simple read/write workloads. These benchmarks often consider scalability to be only the system's ability to use additional hardware resources (scale-out) \cite{b15}. Nevertheless, there is still insufficient consideration of the logical alignment between the data topology and the database model. As noted in \cite{b16}, performance bottlenecks in Big Data systems are often not caused by  hardware limitations, but by an impedance mismatch between the application's query patterns and the underlying storage logic.
Within this setting, recent literature has attempted to bridge the gap between Document and Graph models through Multi-model databases or Polyglot Persistence architectures \cite{b17}. While these approaches theoretically offer the best of both worlds, they introduce significant complexity with respect to data consistency and synchronization \cite{b18}. Furthermore, benchmarks on these hybrid systems often analyze the two models in isolation, testing document retrieval separately from graph traversals, without evaluating the middle ground scenarios where data requires both rich entity descriptions and deep structural navigation \cite{b19}.

Several works have addressed the specific optimization of graph queries on document stores or the storage of heavy attributes on graph nodes \cite{b20}. For example, some Document-based DBMSs frequently employ application-side denormalization techniques to simulate relationships, but these methods shift the computational burden to the application layer, leading to increased latency during complex join operations \cite{b21}. Similarly, loading heavy JSON-like documents into GraphDB nodes can reduce traversal speed due to increased I/O overhead \cite{b22}. Consequently, these optimization attempts often result in trade-offs that are valid for specific edge cases but which fail to generalize to heterogeneous datasets where the workload is mixed \cite{b31}.

Recent surveys have highlighted that although the trade-offs are well documented, the impact of logical scalability, defined as the ability of a data model to sustain performance growth as query complexity increases, remains underexplored \cite{b23}. Current research often presents the choice between document and graph models as a matter of prioritizing flexibility or connectivity, while overlooking the quantitative effects of this decision on the underlying computing infrastructure. Consequently, computational bottlenecks are often incorrectly attributed to hardware limitations rather than structural inefficiencies, resulting in costly and ineffective scaling strategies \cite{b24}.

%% file: tex/motivation.tex
\section{Motivation} \label{sec:problem_identification}

\begin{figure}[ht!]
\centering
\includegraphics[width=0.75\linewidth]{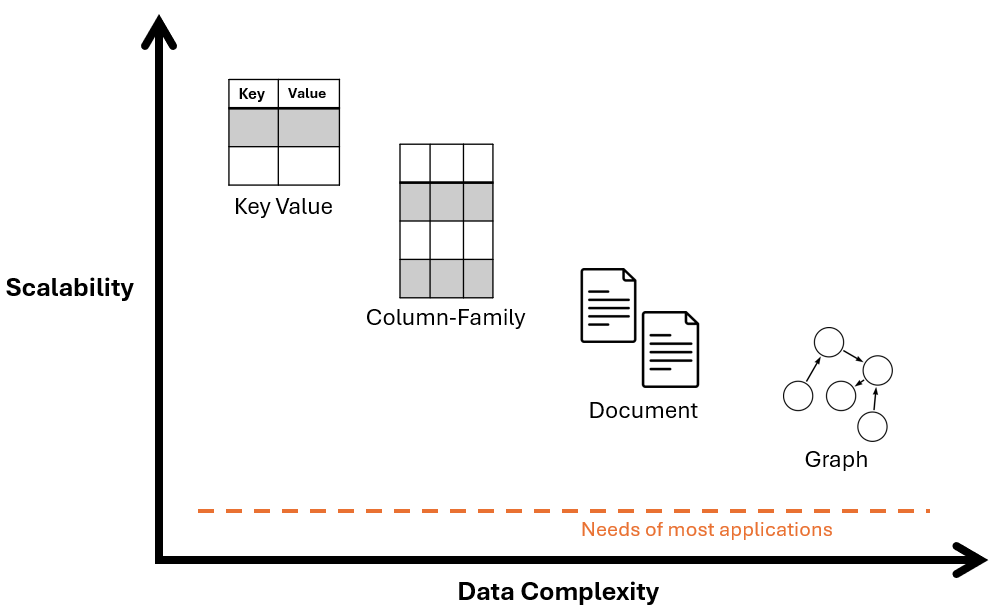}
\caption{Classification of NoSQL paradigms in terms of scalability and complexity.}
\label{fig:complexity_scalability}
\end{figure}

In big data environments, system architectures are organized around the data lifecycle, which includes ingestion, storage, and analysis. The storage layer plays a central role in this process, as it must handle continuous data streams while maintaining fast query performance. Within this ecosystem, different logical models have been developed, each designed to balance scalability and data complexity in specific ways (Fig. \ref{fig:complexity_scalability}).

In this context, Document and Graph paradigms represent the two most significant approaches in the current landscape.  Document-oriented databases are designed for workloads involving high write rates and loosely structured data. They perform best when operations focus on self-contained records, prioritizing horizontal scalability. In contrast, Graph databases are designed to explicitly represent relationships, enforcing consistency to preserve the data structure. These systems are optimized for workloads where value is derived from traversing networks of interconnected entities rather than processing isolated objects.

However, real-world applications rarely fit neatly into these separate categories.  A significant "middle ground" exists where workloads require both the flexible retrieval of documents and the execution of complex traversals to uncover patterns. To address this, the market has adopted Hybrid (or Multi-Model) systems that combine document storage with graph capabilities. While this approach reduces the need to maintain separate databases, it raises questions regarding efficiency and whether a unified model can match the performance of specialized engines.

A prime example of this necessity is found in Value Added Tax (VAT) Fraud detection, specifically for identifying "carousel fraud" and "missing trader" schemes. These fraudulent activities represent a significant challenge for Big Data architectures because they shift the system's requirements from simple data persistence to the identification of suspicious topological patterns across millions of interconnected entities \cite{b27}.
Missing Trader Intra-Community (MTIC) Fraud serves as the foundational element of these schemes, where a "missing trader" imports goods from another EU member state VAT-free, sells them in the local market with VAT included, and then vanishes without remitting the collected tax to the government \cite{b28}. This scenario requires a storage layer capable of identifying companies that exhibit high-volume transactional ingestion but lack a stable historical or relational footprint. Carousel Fraud is a more sophisticated evolution of this scheme, involving a circular chain of transactions where goods or services are traded multiple times across different countries to generate several fraudulent VAT refund claims \cite{b29}. Such a scenario demands a storage layer capable of managing  millions of  nodes and millions of relationships, requiring the system to trace long flows of goods or money across multiple distinct entities (e.g., 4-hop traversals). 

The scaling requirements for these applications are twofold: the engine must sustain high-volume transactional ingestion while simultaneously minimizing the impedance mismatch during deep structural pattern matching.  A common issue in this domain is the tendency to view scalability only as an infrastructural challenge. Standard strategies often focus on resolving bottlenecks by adding hardware resources, either by increasing processor power (vertical scaling) or by adding nodes to a cluster (horizontal scaling). However, this perspective overlooks the logical scalability of the data model, as it determines the system's scaling potential.
Consequently, physical scalability is ineffective without logical scalability.  Therefore, efficient big data analysis requires shifting focus from simply adding resources to selecting the correct logical model.

%% file: tex/material_and_methods.tex
\section{Material and Methods}\label{sec:methods}
To evaluate the impact of logical scalability on the performance of Big Data systems, we designed a comparative framework that considers the logical model as the primary factor responsible for the system’s scalability. Unlike standard benchmarks that focus on random read/write operations, we categorize queries based on their structural complexity to measure how the logical model scales with the shape of the question. This section outlines the system architecture, the dataset's characteristics and the experimental methodology employed to benchmark Document-oriented, Graph-based and Multi-model solutions.

\subsection{System Architecture}

The architecture deploys three parallel environments that share identical hardware resources (CPU cores, RAM and I/O bandwidth). The experimental setup comprises three distinct database paradigms:

\begin{itemize}
    \item \textbf{Document Store (MongoDB):} Selected as the industry standard for the Availability/Partition-tolerance (AP) paradigm. It stores data as JSON-like documents, prioritizing write throughput and schema flexibility. In this experiment, relationships are modeled via referencing, requiring the use of the aggregation pipeline (specifically the \texttt{\$lookup} stage) to simulate joins.
    \item \textbf{Graph Database (Neo4j):} Selected as the leading solution for the Consistency/Partition-tolerance (CP) paradigm. It utilizes \textit{index-free adjacency} to store nodes and relationships natively, optimizing traversal operations over unconnected data blobs using the Cypher query language.
    \item \textbf{Multi-Model Database (ArangoDB):} Selected to represent the hybrid approach. ArangoDB allows for valid comparisons between document storage and graph traversal within a single engine. It utilizes AQL (ArangoDB Query Language) to perform both filtering and graph traversals (using \texttt{OUTBOUND} directionality) on the same dataset.
\end{itemize}

All database interactions are orchestrated via a Python-based control module using the official drivers: \texttt{pymongo} for MongoDB, the \texttt{neo4j} Bolt driver for Neo4j, and \texttt{python-arango} for ArangoDB.

In addition to the single-node experiments, further evaluations were conducted in a distributed configuration. Each DBMS was deployed in a clustered environment composed of four virtual machines. Each VM runs Ubuntu Server LTS and is provisioned with 8 vCPUs and 16 GB of RAM. The distributed setup enables the assessment of query execution behavior under horizontal scaling conditions, allowing the analysis of performance, resource utilization, and inter-node communication overhead in a multi-node cluster scenario. The cluster size was limited to four virtual machines due to licensing constraints associated with the Neo4j Community Edition, which does not support larger production-grade clustered deployments. Extending the cluster beyond this configuration would have required the adoption of the Neo4j Enterprise Edition. Therefore, to ensure methodological consistency and maintain comparable experimental conditions across all evaluated DBMSs, the distributed environment was standardized to a four-node cluster.

\subsection{Dataset Preparation and Database Selection}
\begin{figure}[ht!]
    \centering
    \includegraphics[width=0.45\textwidth]{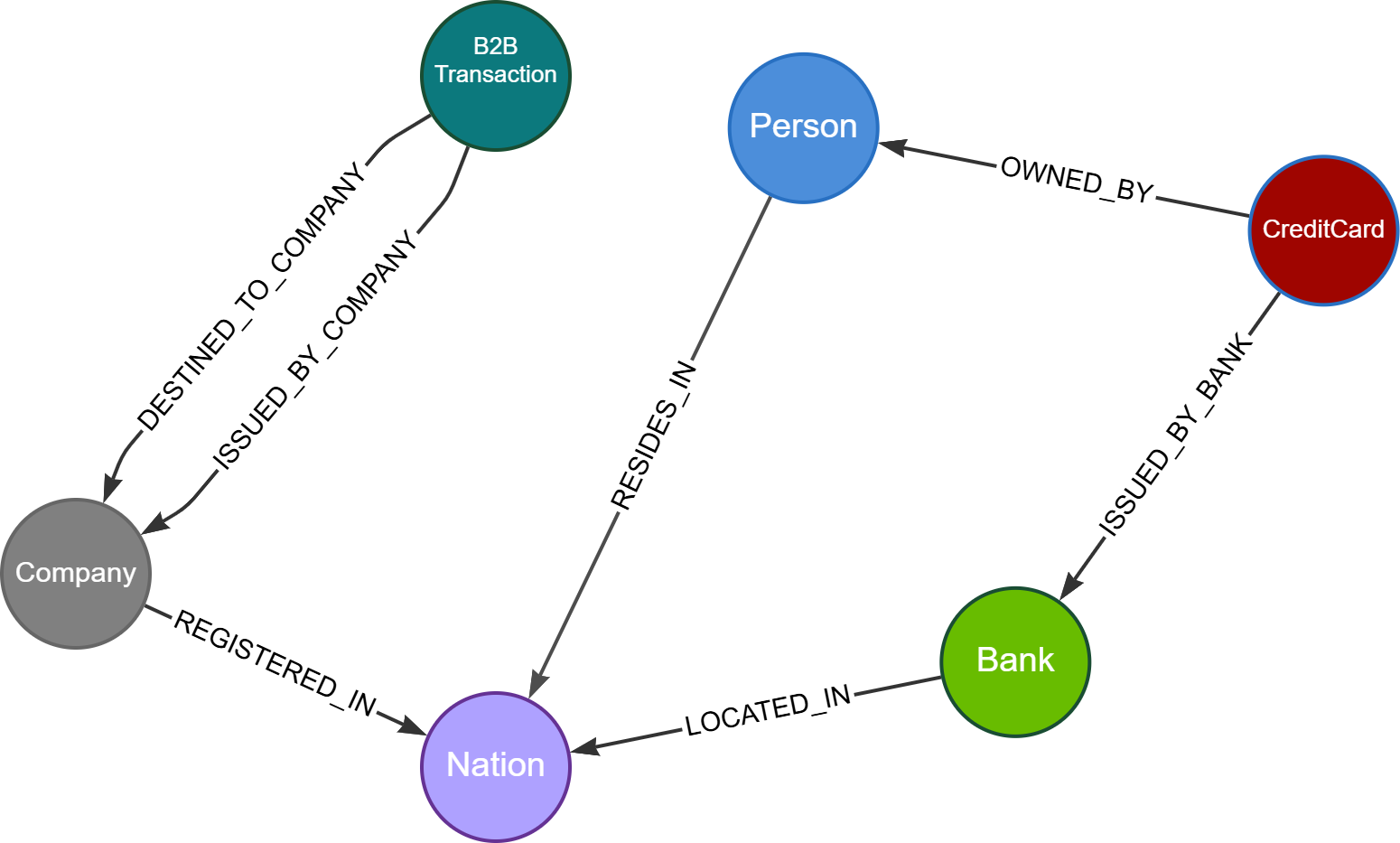}
    \caption{Schema of the VAT Fraud Detection.}
    \label{fig:fraud_schema}
\end{figure}

The experimental evaluation was conducted using a large-scale synthetic dataset generated to simulate realistic Value Added Tax (VAT) fraud scenarios, specifically focusing on "carousel fraud" patterns and "missing trader" structures.  The dataset was formalized into a Property Graph, where connections are explicitly modeled as relationships rather than foreign keys. The logical schema (Fig. \ref{fig:fraud_schema}), consists of six distinct node labels: Company, Nation, B2B Transaction, Person, Credit Card and Bank.

The study aims to compare the performance of three distinct NoSQL paradigms in identifying complex fraud patterns:
\begin{itemize}
    \item \textbf{MongoDB}: Representing the document-oriented paradigm, storing data in semi-structured BSON collections.
    \item \textbf{ArangoDB}: Representing the multi-model paradigm, utilized here to leverage its native graph capabilities alongside document storage.
    \item \textbf{Neo4j}: Representing the native graph paradigm, focusing on index-free adjacency for deep traversal performance.
\end{itemize}
These entities were modeled as Nodes and Relationships in Neo4j and ArangoDB, and as normalized or semi-normalized Collections in MongoDB.

\subsection{Scaling Strategy and Dataset Dimensions}
To evaluate the scalability of the three systems, a rigorous scaling strategy was adopted. A baseline synthetic dataset, representing the "100\% Profile," was generated to stress-test the query engines with high-volume transactional data. 
The full-scale baseline dataset (100\% profile) is characterized by significant cardinality, comprising 190 Nations, 500 Banks, 70,000 Companies, 150,000 People, 100,000 Credit Cards, and 1.500.000 B2B Transactions.

From this baseline, three subsets were derived to analyze performance trends across MongoDB, ArangoDB, and Neo4j:
\begin{itemize}
    \item \textbf{100\% Profile}: The full dataset, representing the maximum load scenario.
    \item \textbf{75\% Profile}: A high-load scenario to test near-maximum capacity performance.
    \item \textbf{50\% Profile}: A medium-load scenario to observe linear scalability.
    \item \textbf{25\% Profile}: A low-load scenario used to establish baseline execution times.
\end{itemize}

This incremental approach allows for the assessment of how execution times degrade or maintain stability as the volume of data increases across the different architectural implementations.
To prepare the data for heterogeneous systems, the graph was first exported into intermediate CSV files, with nodes categorized by label and relationships by type. Each CSV record contains the element’s unique Neo4j identifier and a JSON-serialized representation of its properties, which handles nested dictionaries and lists. Two distinct sets of JSON documents were generated from these intermediate files to match the target data models:

\begin{itemize}
    \item \textbf{For MongoDB:} Node documents preserve the original Neo4j ID and encapsulate all attributes within a 'properties' field containing the de-serialized JSON object. Relationships are explicitly modeled as documents containing \texttt{id}, \texttt{from}, \texttt{to}, and \texttt{properties}.
    \item \textbf{For ArangoDB:} A global lookup table was generated to map Neo4j internal IDs to their corresponding labels. This enabled the construction of valid \texttt{\_key}, \texttt{\_from}, and \texttt{\_to} fields in the required format \verb|<collection>/<_key>|, while preserving all original information under a unified \texttt{properties} field.
\end{itemize}

\subsection{System Ingestion}
The ingestion layer uses the respective native drivers of the target data stores to transfer the prepared JSON representations, thereby ensuring experimental fairness. For \textbf{MongoDB}, a custom Python client establishes a connection to the database and iterates over each collection of node and relationship documents to perform insertions.
In the \textbf{ArangoDB} pipeline, the ingestion script leverages the ArangoDB Python driver to manage the hybrid structure. Node documents are loaded into their respective vertex collections, while edge documents are inserted into the designated edge collections, utilizing the pre-formatted \texttt{\_from} and \texttt{\_to} fields to immediately establish topological connections.

%% file: tex/results.tex
\section{Results}

Using a custom schema analysis algorithm, we automatically generate four classes of 
semantically equivalent queries for each database. To assess \textit{Logical Scalability} 
while accounting for driver overhead and caching mechanisms, we implemented a rigorous 
execution protocol divided into two phases for every query and dataset size combination:

\begin{itemize}
    \item \textbf{Cold Runs (Latency Focus):} 
    We perform 31 iterations where a new database connection is established and strictly 
    closed for every single query execution. This measures the total round-trip time, 
    including the driver handshake and the system's performance on \textit{cold} data 
    (uncached).
    
    \item \textbf{Warm Runs (Throughput Focus):} 
    We perform 30 iterations using a persistent connection. This measures the pure query 
    execution time, allowing the database internal caches (page cache, query plan cache) 
    to warm up and stabilize.
\end{itemize}

\textbf{Metrics:} The primary metric recorded is \textbf{Execution Time (ms)}. Results 
are aggregated to calculate the mean execution time and 95\% confidence intervals. By 
keeping the hardware constant and varying both the dataset size ($1x$ to $4x$) and query 
complexity (Q1 to Q4), we provide empirical evidence of where \textit{Hardware 
Scalability} fails and Logical Scalability becomes the decisive factor.

\subsection{Q1 - Complex Scan (Multi-Attribute Filtering)}
This query performs a scan on the B2BTransaction collection, applying concurrent filters 
on numerical ranges (checking if total\_eur is between 100 and 500,000 and 
VAT $\geq$ 0.10) and temporal windows between 2020 and 2024. This tests the 
engine's ability to handle compound indexing and selective retrieval on a single 
collection without graph traversal.

\begin{figure}[ht]
\centering
\fbox{
    \begin{minipage}{0.95\columnwidth}
        \centering
        \textbf{Query 1}\\[0.2cm]
        \begin{minipage}{0.48\columnwidth}
            \centering
            \includegraphics[width=\linewidth]{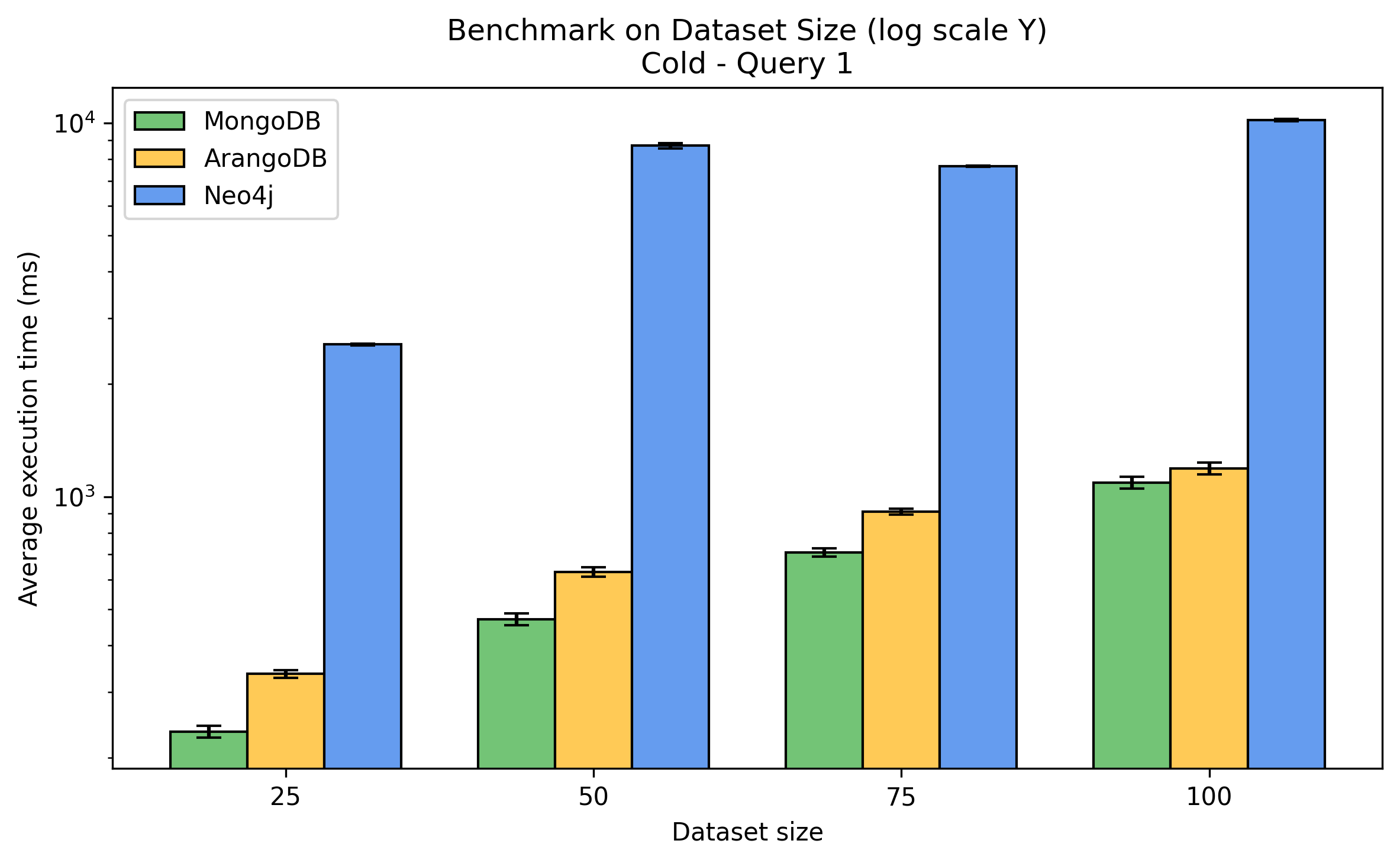}\\
            \footnotesize Cold
        \end{minipage}
        \hfill
        \begin{minipage}{0.48\columnwidth}
            \centering
            \includegraphics[width=\linewidth]{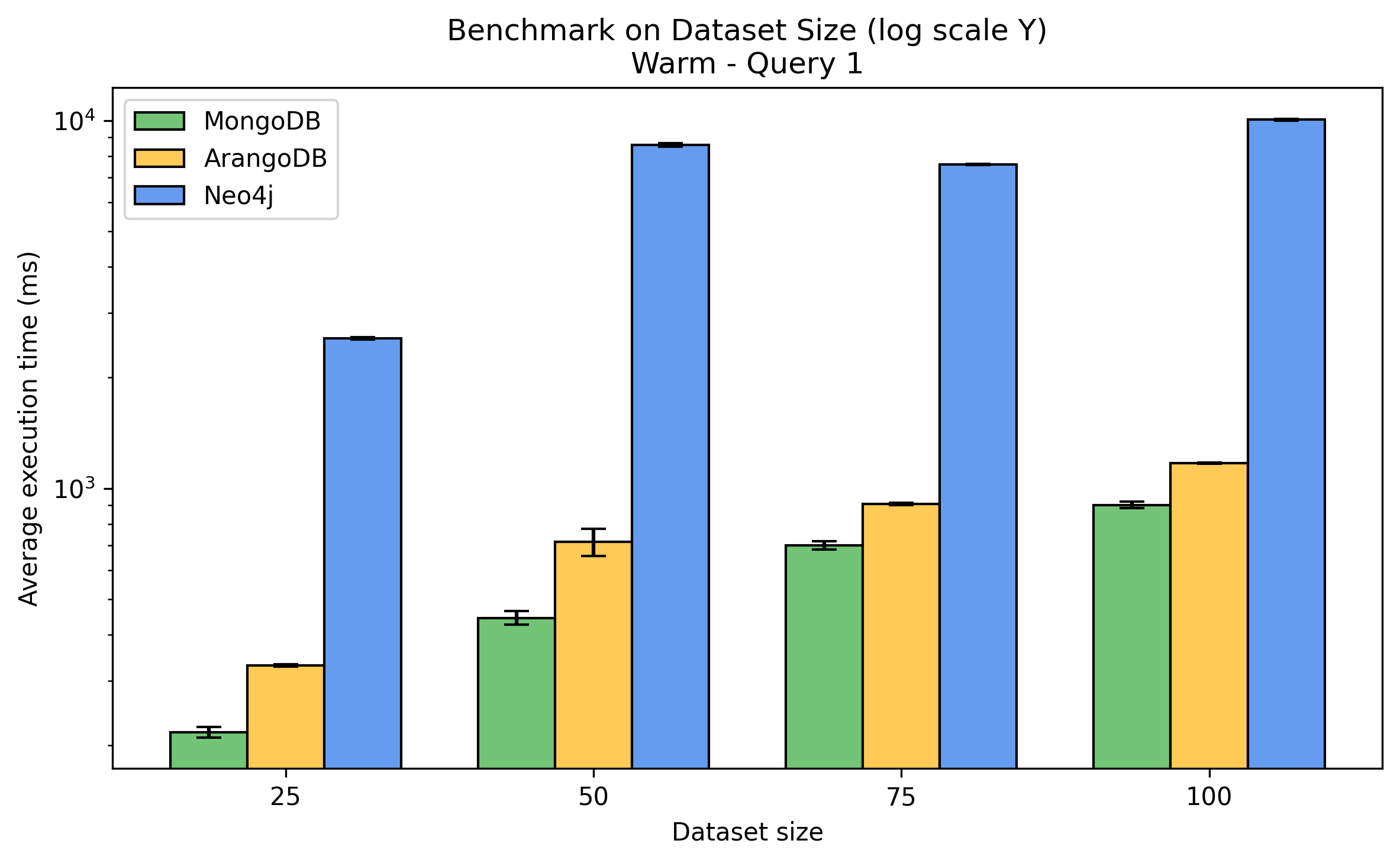}\\
            \footnotesize Warm
        \end{minipage}
    \end{minipage}
}
\caption{Query 1 Results.}
\end{figure}

\begin{figure}[ht]
\centering
\fbox{
    \begin{minipage}{0.95\columnwidth}
        \centering
        \textbf{Query 1 (Distributed Cluster with 4VM)}\\[0.2cm]
        \begin{minipage}{0.48\columnwidth}
            \centering
            \includegraphics[width=\linewidth]{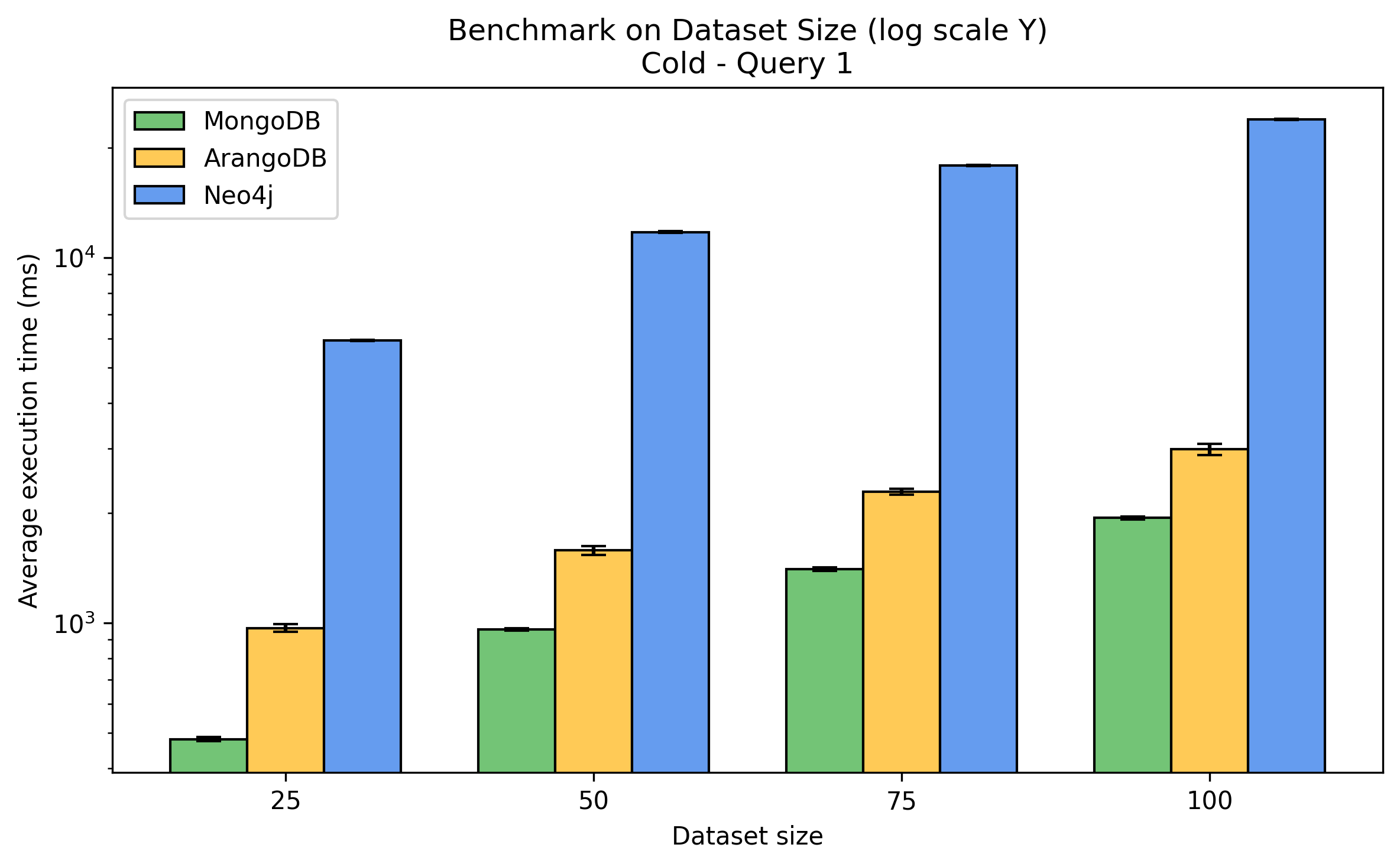}\\
            \footnotesize Cold
        \end{minipage}
        \hfill
        \begin{minipage}{0.48\columnwidth}
            \centering
            \includegraphics[width=\linewidth]{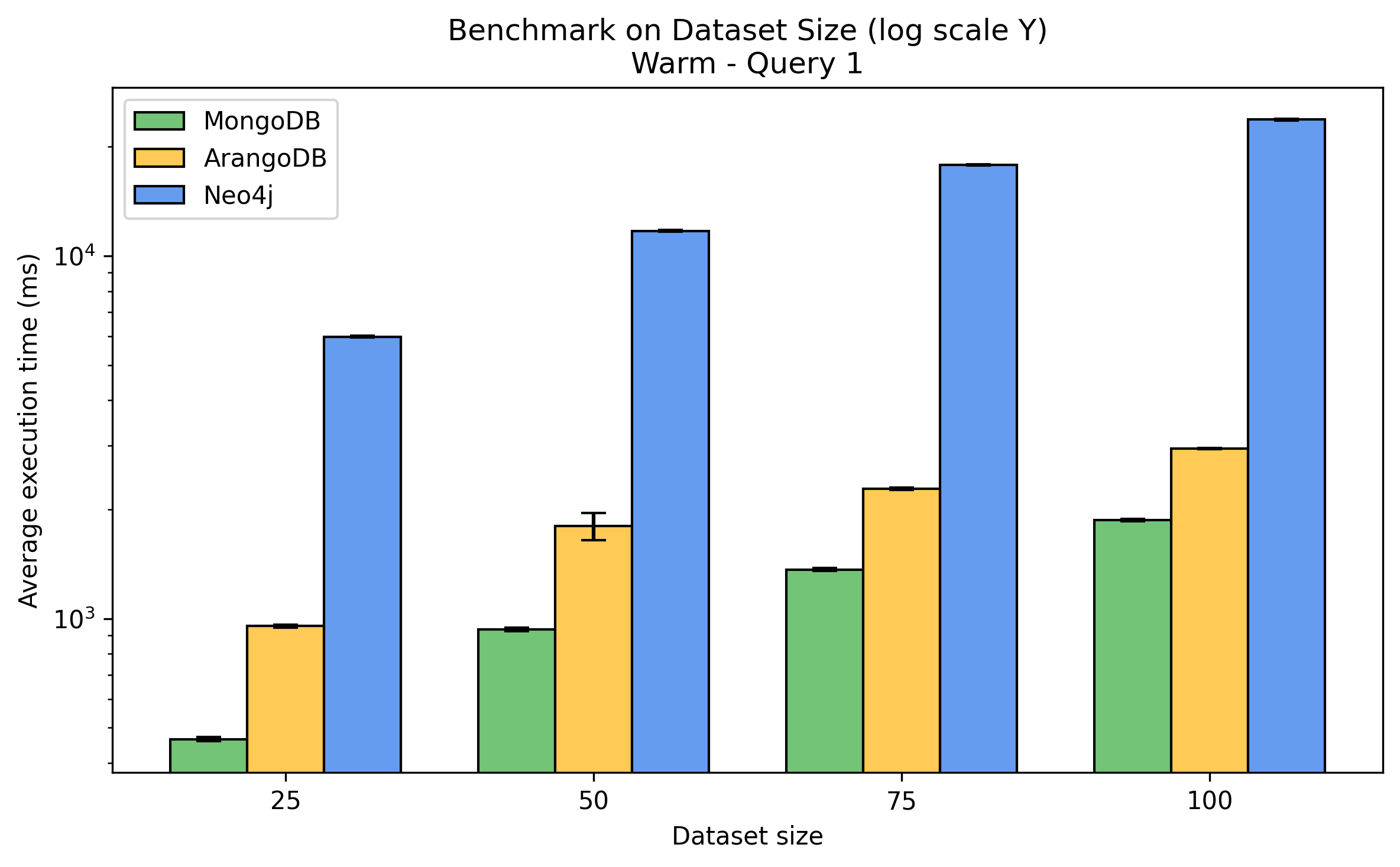}\\
            \footnotesize Warm
        \end{minipage}
    \end{minipage}
}
\caption{Query 1 Results in Distributed Cluster.}
\end{figure}

Q1 results show that MongoDB achieves the lowest execution times across all dataset profiles, followed by ArangoDB, while Neo4j is consistently slower by approximately one order of magnitude in both cold and warm conditions. This reflects a structural misalignment between Neo4j's index-free adjacency mechanism and the nature of this query, which requires only attribute-based filtering on isolated node collections. MongoDB and ArangoDB use compound indexes directly on document fields, making selective retrieval efficient at any scale. All three systems exhibit linear growth in execution time from 25\% to 100\%, confirming that performance degradation here is driven by data volume rather than structural complexity.

\subsection{Q2 - 1-Hop Join (Company-Transaction Aggregation)}
The second query executes a traversal or join operation connecting every Company node to its 
issued B2BTransactions. This represents a standard master-detail retrieval 
(or ``star schema'' join), measuring the efficiency of document embedding/lookups (in 
MongoDB) versus neighbor expansion (in Neo4j/ArangoDB) without deep path navigation.

\begin{figure}[ht]
\centering
\fbox{
    \begin{minipage}{0.95\columnwidth}
        \centering
        \textbf{Query 2}\\[0.2cm]
        \begin{minipage}{0.48\columnwidth}
            \centering
            \includegraphics[width=\linewidth]{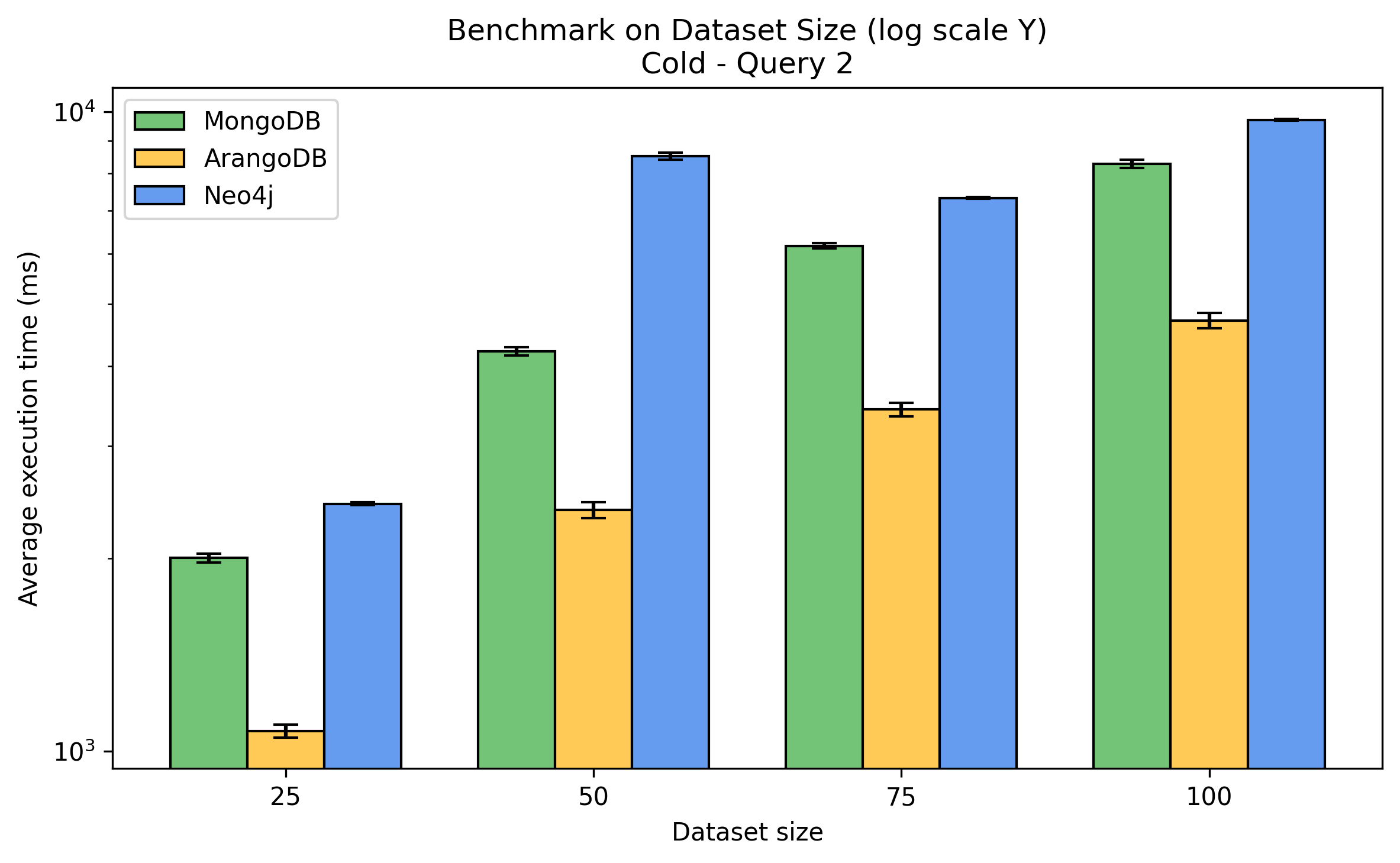}\\
            \footnotesize Cold
        \end{minipage}
        \hfill
        \begin{minipage}{0.48\columnwidth}
            \centering
            \includegraphics[width=\linewidth]{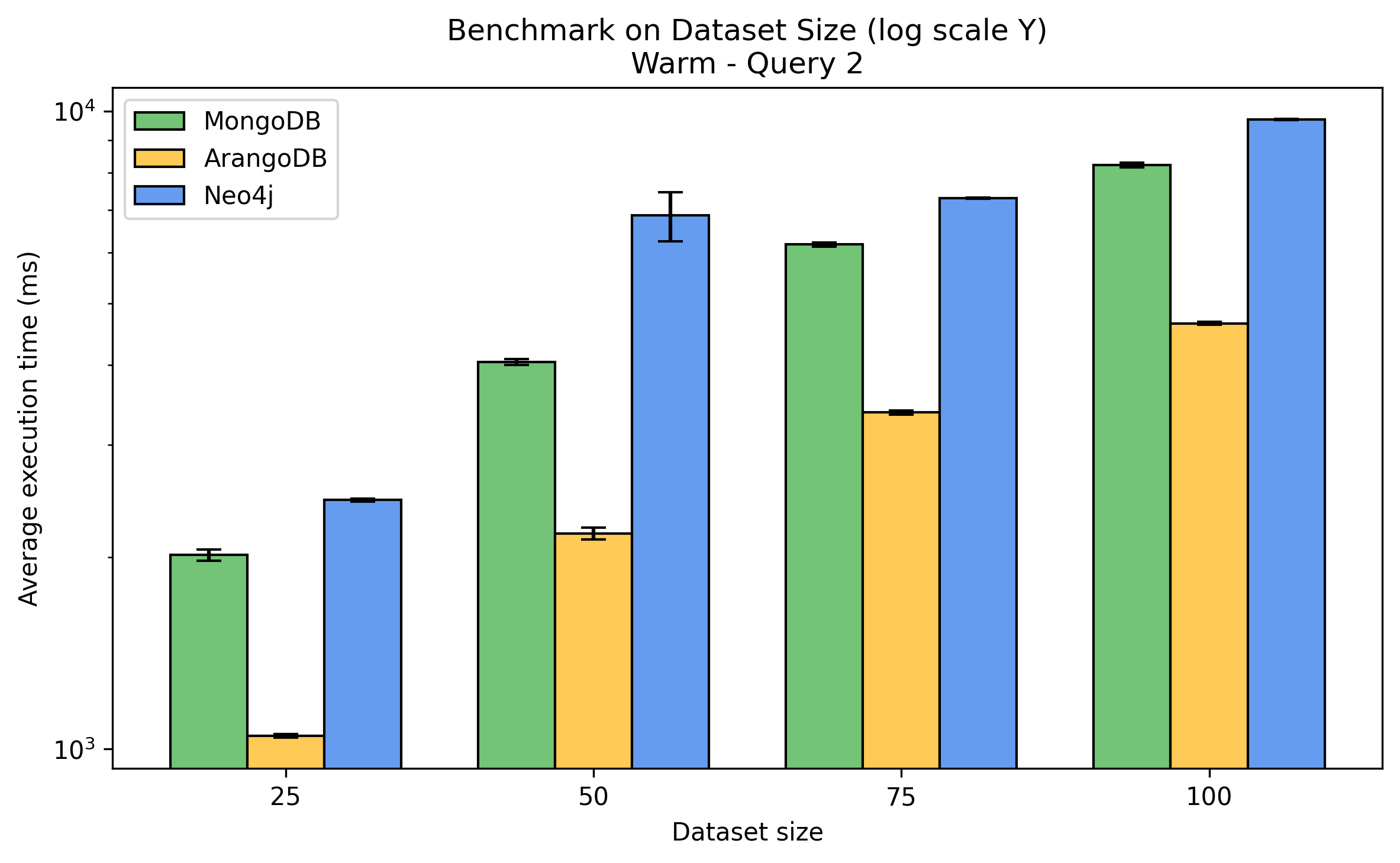}\\
            \footnotesize Warm
        \end{minipage}
    \end{minipage}
}
\caption{Query 2 Results.}
\end{figure}

\begin{figure}[ht]
\centering
\fbox{
    \begin{minipage}{0.95\columnwidth}
        \centering
        \textbf{Query 2 (Distributed Cluster with 4VM)}\\[0.2cm]
        \begin{minipage}{0.48\columnwidth}
            \centering
            \includegraphics[width=\linewidth]{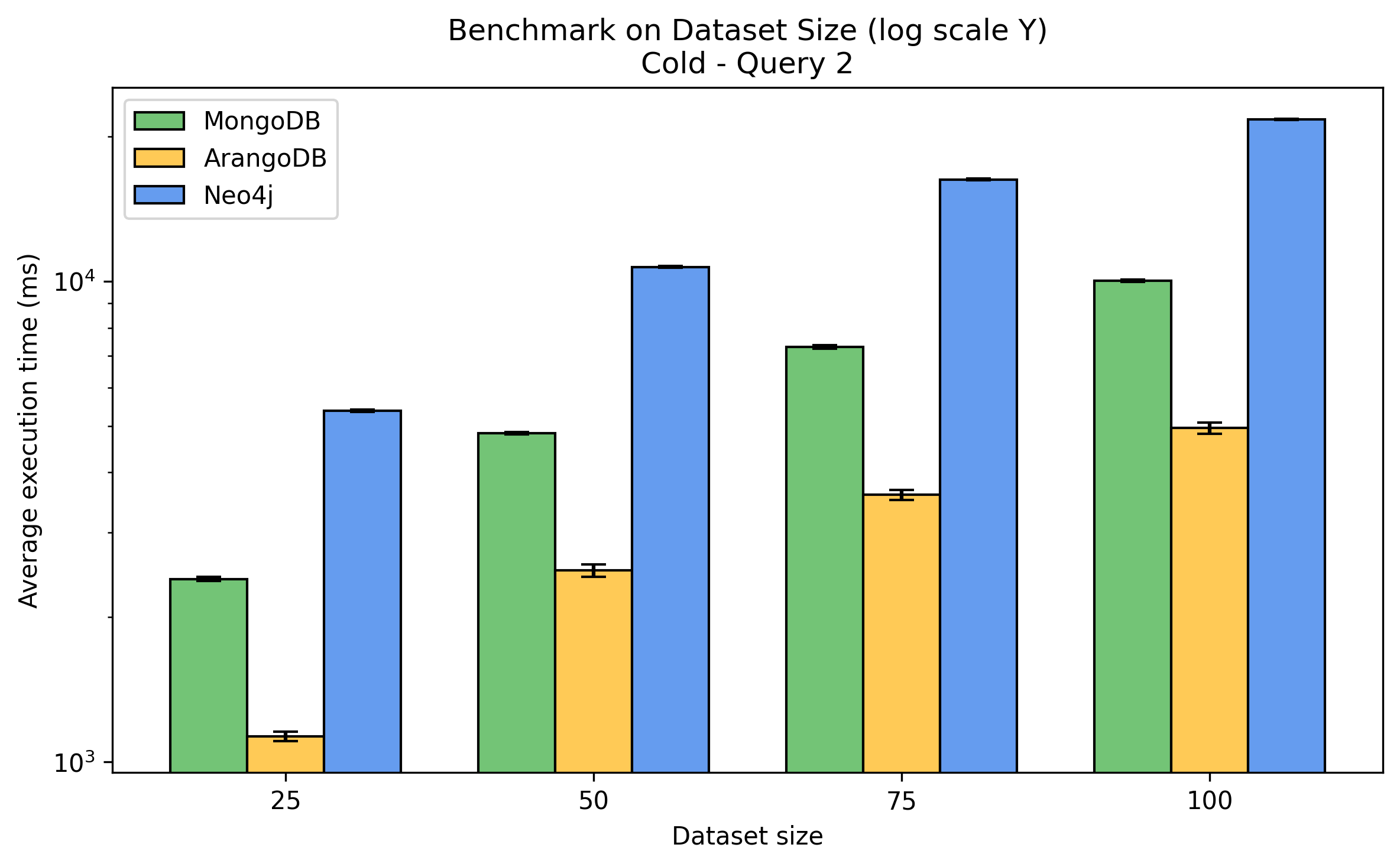}\\
            \footnotesize Cold
        \end{minipage}
        \hfill
        \begin{minipage}{0.48\columnwidth}
            \centering
            \includegraphics[width=\linewidth]{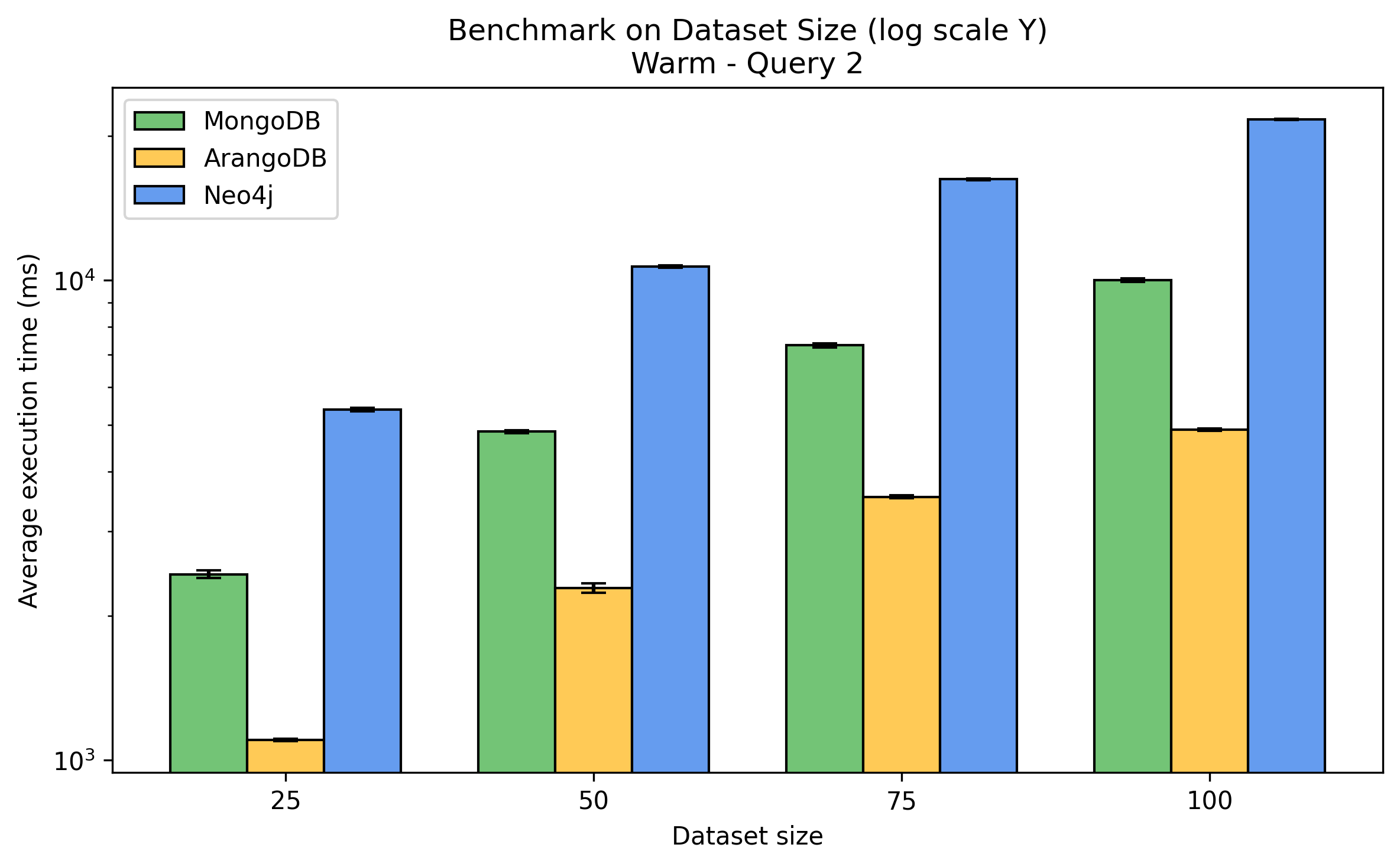}\\
            \footnotesize Warm
        \end{minipage}
    \end{minipage}
}
\caption{Query 2 Results in Distributed Cluster.}
\end{figure}

Q2 results reveal a shift in relative performance as the query introduces a relational dimension. ArangoDB achieves the lowest execution times across all profiles, while MongoDB degrades more steeply with scale due to the \$lookup stage in the aggregation pipeline, which resolves inter-document references at query time with a cost proportional to the number of Company-Transaction pairs. Neo4j, starting from latencies comparable to ArangoDB at 25\%, exceeds MongoDB at the 100\% profile, as the global nature of the join, retrieving all transactions for every company, results in proportionally higher latency despite direct edge traversal.

\subsection{Q3 - Short Chain Analysis (Unbounded 2-Hop)}
The third query identifies direct commercial relationships between two companies mediated by a 
transaction (\textit{Company A} $\leftarrow$ \textit{Transaction} $\rightarrow$ 
\textit{Company B}). This global 2-hop traversal runs without a LIMIT clause, 
stress-testing the system's throughput and memory management when retrieving all 
direct B2B trading partners across the entire dataset.

\begin{figure}[ht]
\centering
\fbox{
    \begin{minipage}{0.95\columnwidth}
        \centering
        \textbf{Query 3}\\[0.2cm]
        \begin{minipage}{0.48\columnwidth}
            \centering
            \includegraphics[width=\linewidth]{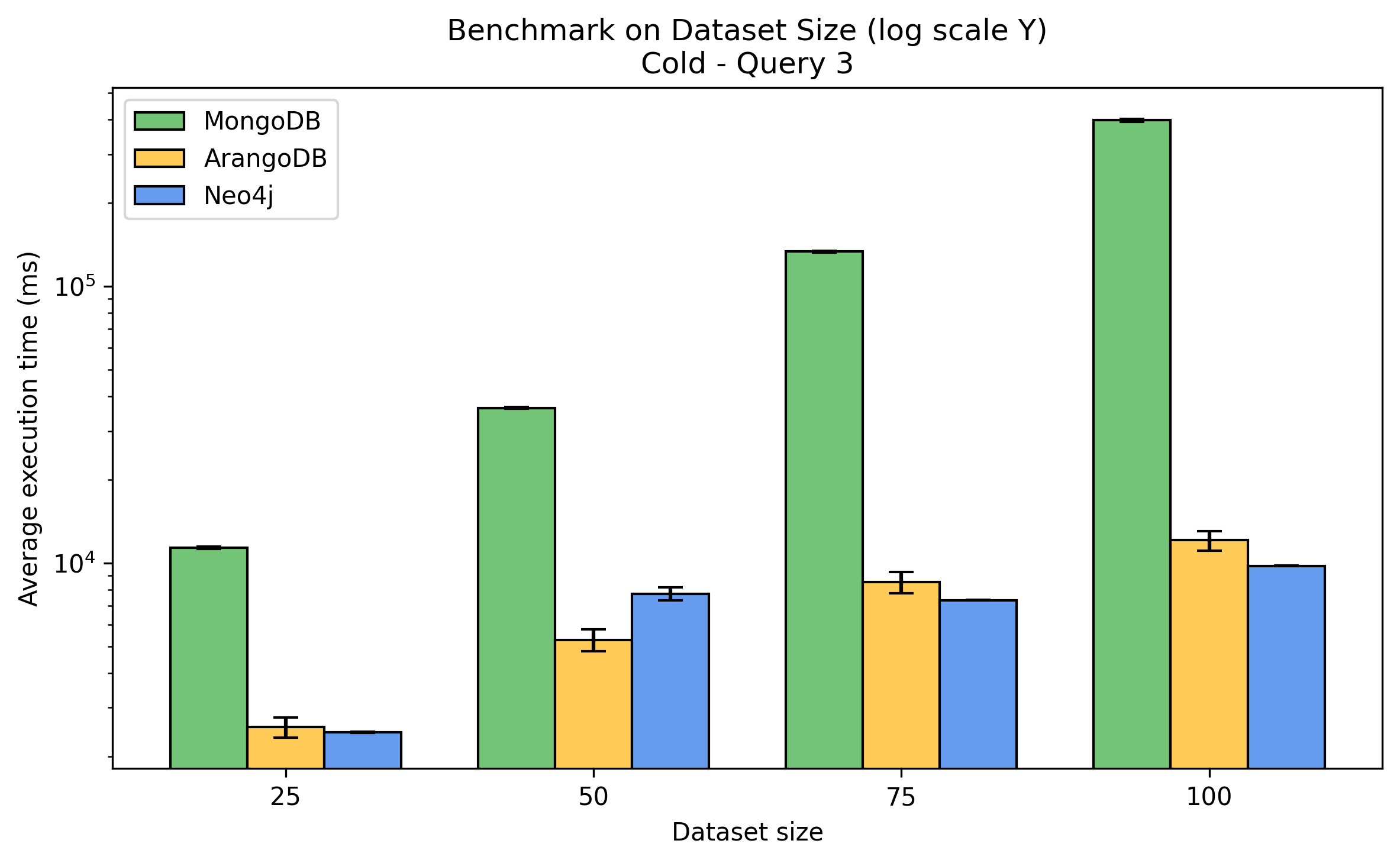}\\
            \footnotesize Cold
        \end{minipage}
        \hfill
        \begin{minipage}{0.48\columnwidth}
            \centering
            \includegraphics[width=\linewidth]{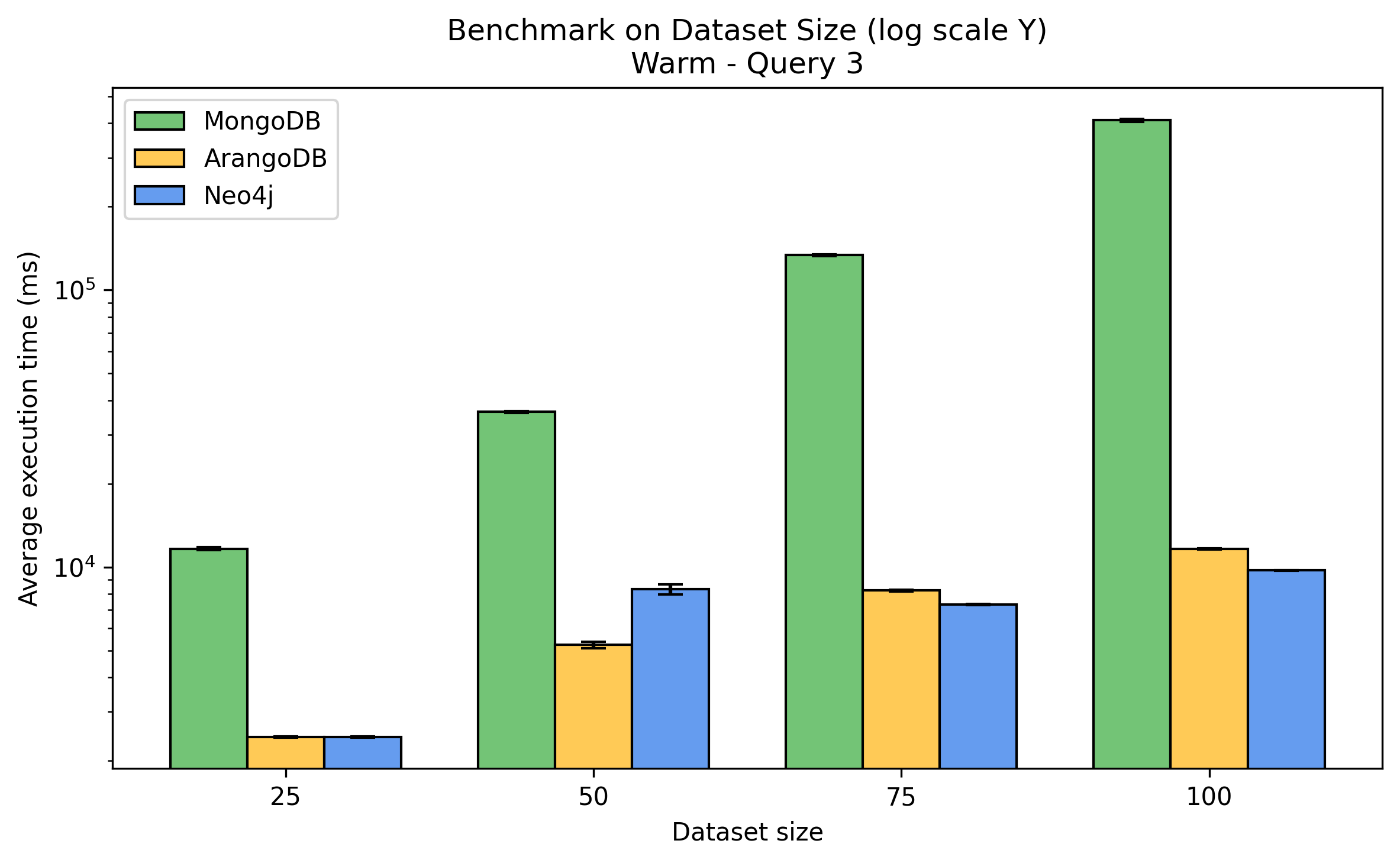}\\
            \footnotesize Warm
        \end{minipage}
    \end{minipage}
}
\caption{Query 3 Results.}
\end{figure}

\begin{figure}[ht]
\centering
\fbox{
    \begin{minipage}{0.95\columnwidth}
        \centering
        \textbf{Query 3 (Distributed Cluster with 4VM)}\\[0.2cm]
        \begin{minipage}{0.48\columnwidth}
            \centering
            \includegraphics[width=\linewidth]{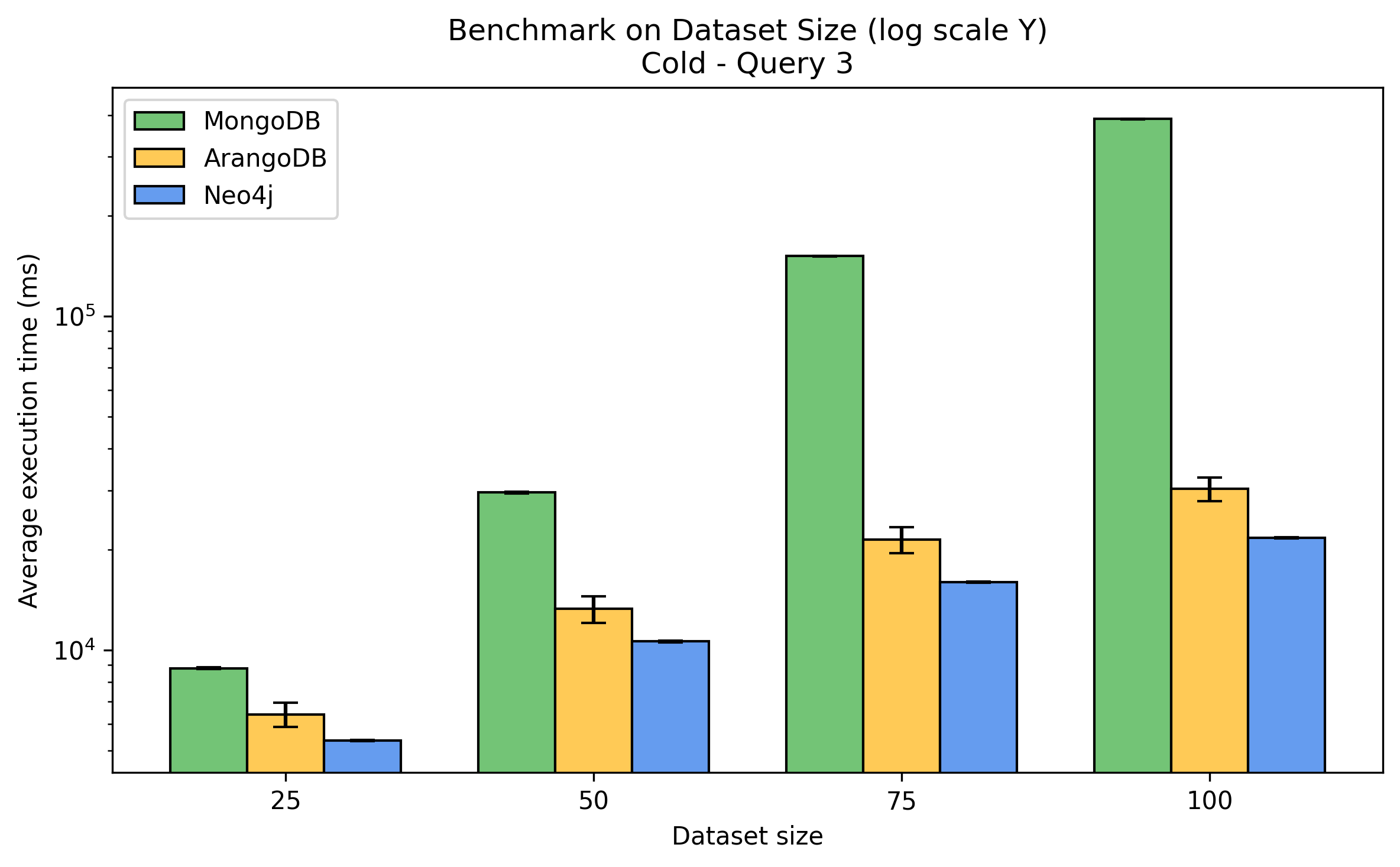}\\
            \footnotesize Cold
        \end{minipage}
        \hfill
        \begin{minipage}{0.48\columnwidth}
            \centering
            \includegraphics[width=\linewidth]{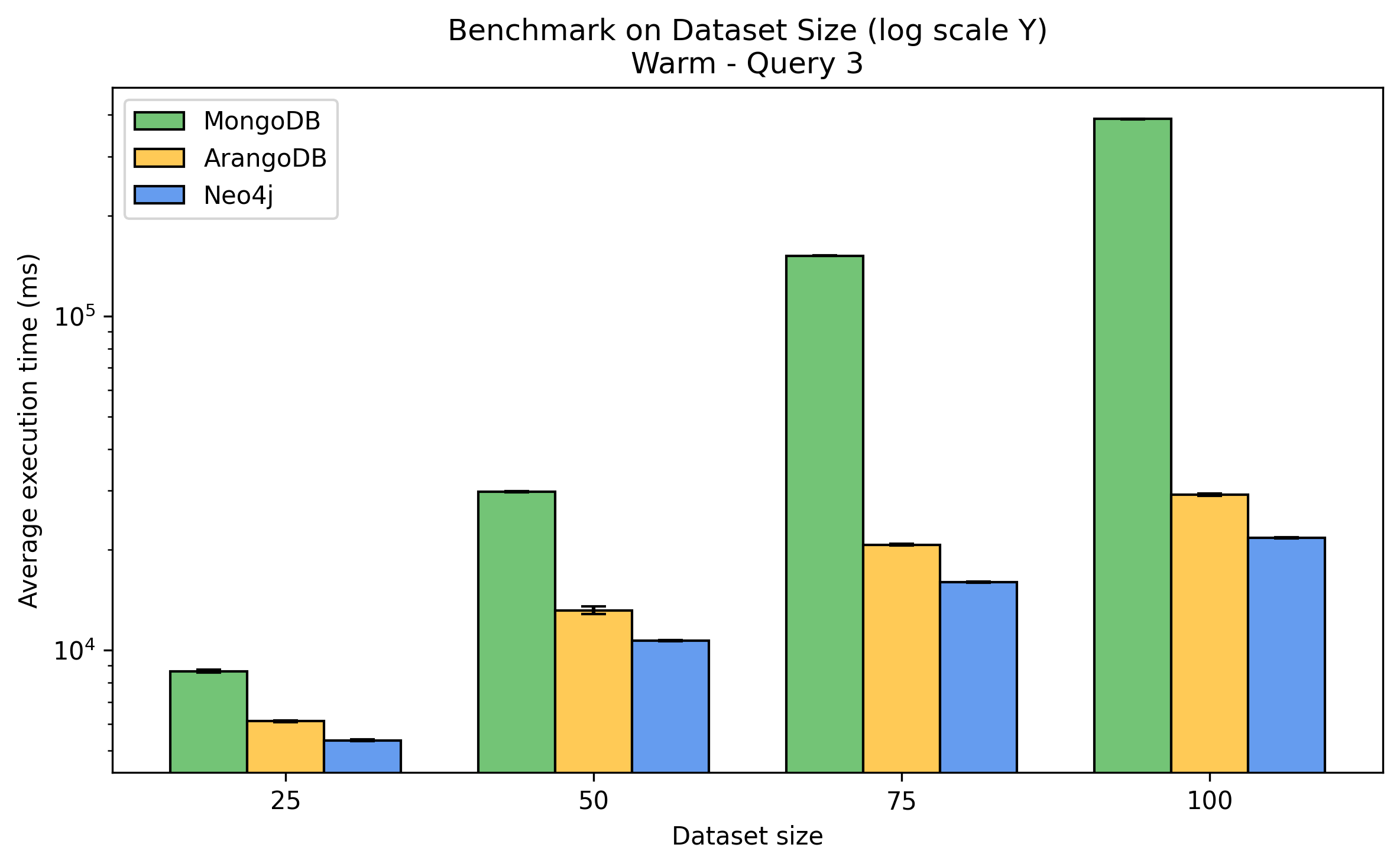}\\
            \footnotesize Warm
        \end{minipage}
    \end{minipage}
}
\caption{Query 3 Results in Distributed Cluster.}
\end{figure}

Q3 results expose a substantial divergence between the document-oriented and graph-based paradigms. MongoDB's execution time grows approximately 30x between the 25\% ($\sim$$10^4$~ms) and 100\% ($\sim$$3 \times 10^5$~ms) profiles, while ArangoDB and Neo4j remain within the $10^4$~ms range with a markedly more contained growth rate, converging to nearly equivalent latencies from the 50\% profile onward. MongoDB's degradation stems from the absence of a LIMIT clause combined with its relational simulation strategy: resolving all Company-to-Company paths requires nested \$lookup operations whose cost grows proportionally to the number of intermediate nodes. Graph-native engines avoid this by navigating edges via direct memory pointers, keeping the per-step traversal cost constant regardless of dataset cardinality.

\subsection{Q4 - Deep Chain Traversal (Limited 4-Hop)}
The last query traces a longer flow of money or goods across three distinct companies 
(\textit{Company A} $\rightarrow$ \textit{Transaction 1} $\rightarrow$ 
\textit{Company B} $\rightarrow$ \textit{Transaction 2} $\rightarrow$ 
\textit{Company C}). This requires a 4-hop traversal (navigating two full transaction 
steps). Crucially, this query applies a \texttt{LIMIT} to the starting nodes (25) and 
results (200) to isolate the latency of deep structural pattern matching from the 
overhead of result serialization.

\begin{figure}[ht]
\centering
\fbox{
    \begin{minipage}{0.95\columnwidth}
        \centering
        \textbf{Query 4}\\[0.2cm]
        \begin{minipage}{0.48\columnwidth}
            \centering
            \includegraphics[width=\linewidth]{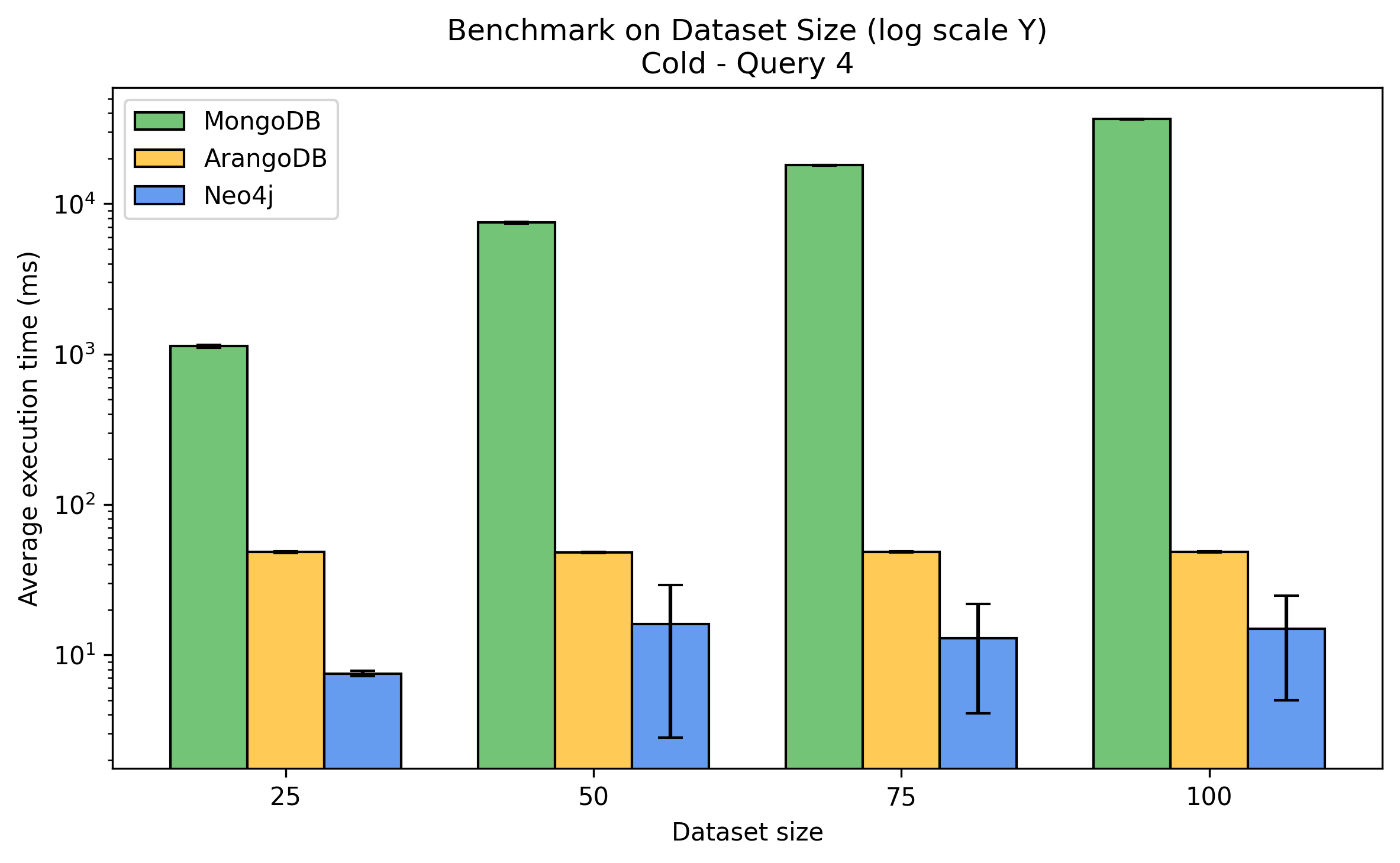}\\
            \footnotesize Cold
        \end{minipage}
        \hfill
        \begin{minipage}{0.48\columnwidth}
            \centering
            \includegraphics[width=\linewidth]{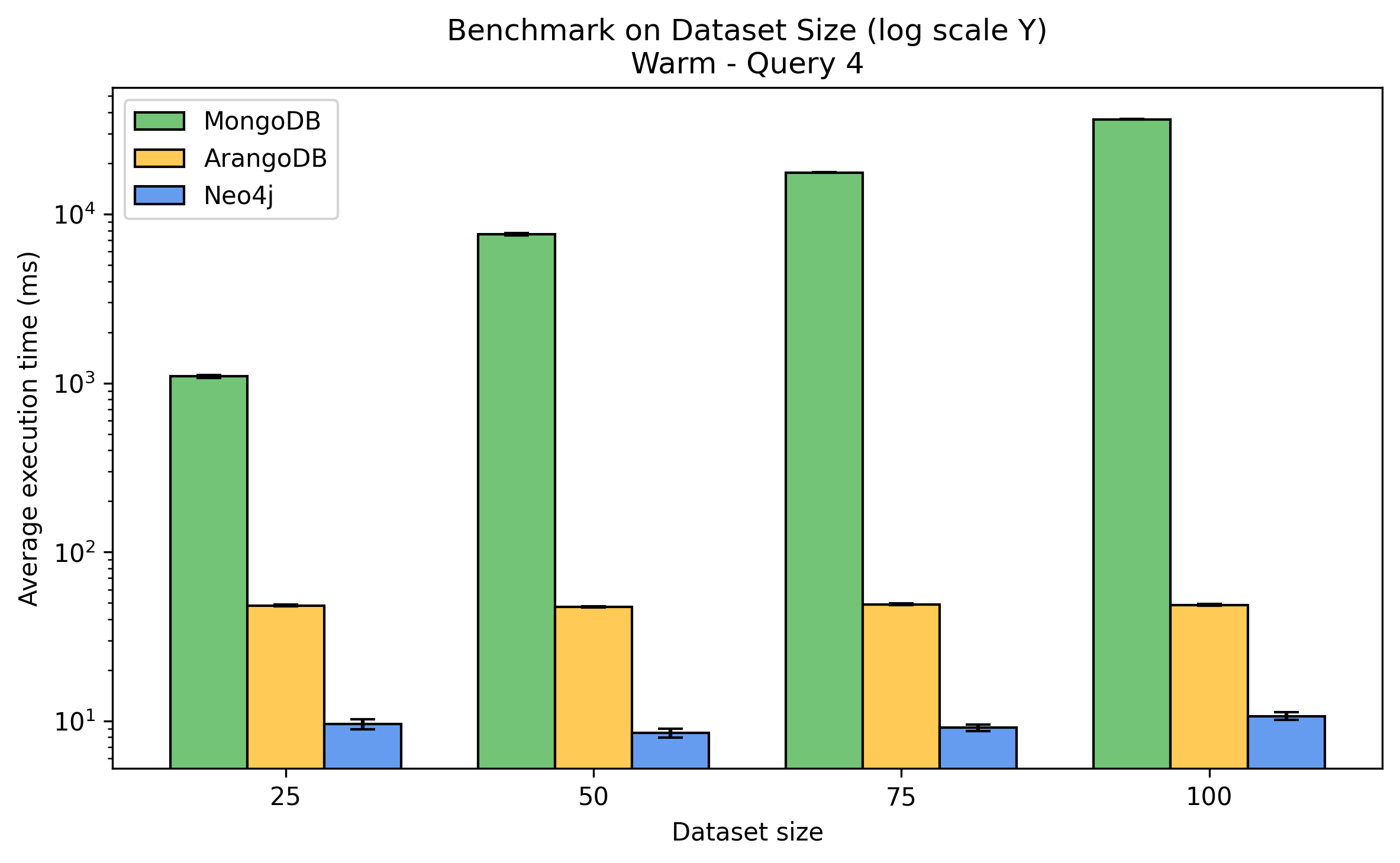}\\
            \footnotesize Warm
        \end{minipage}
    \end{minipage}
}
\caption{Query 4 Results.}
\end{figure}

\begin{figure}[ht]
\centering
\fbox{
    \begin{minipage}{0.95\columnwidth}
        \centering
        \textbf{Query 4 (Distributed Cluster with 4VM)}\\[0.2cm]
        \begin{minipage}{0.48\columnwidth}
            \centering
            \includegraphics[width=\linewidth]{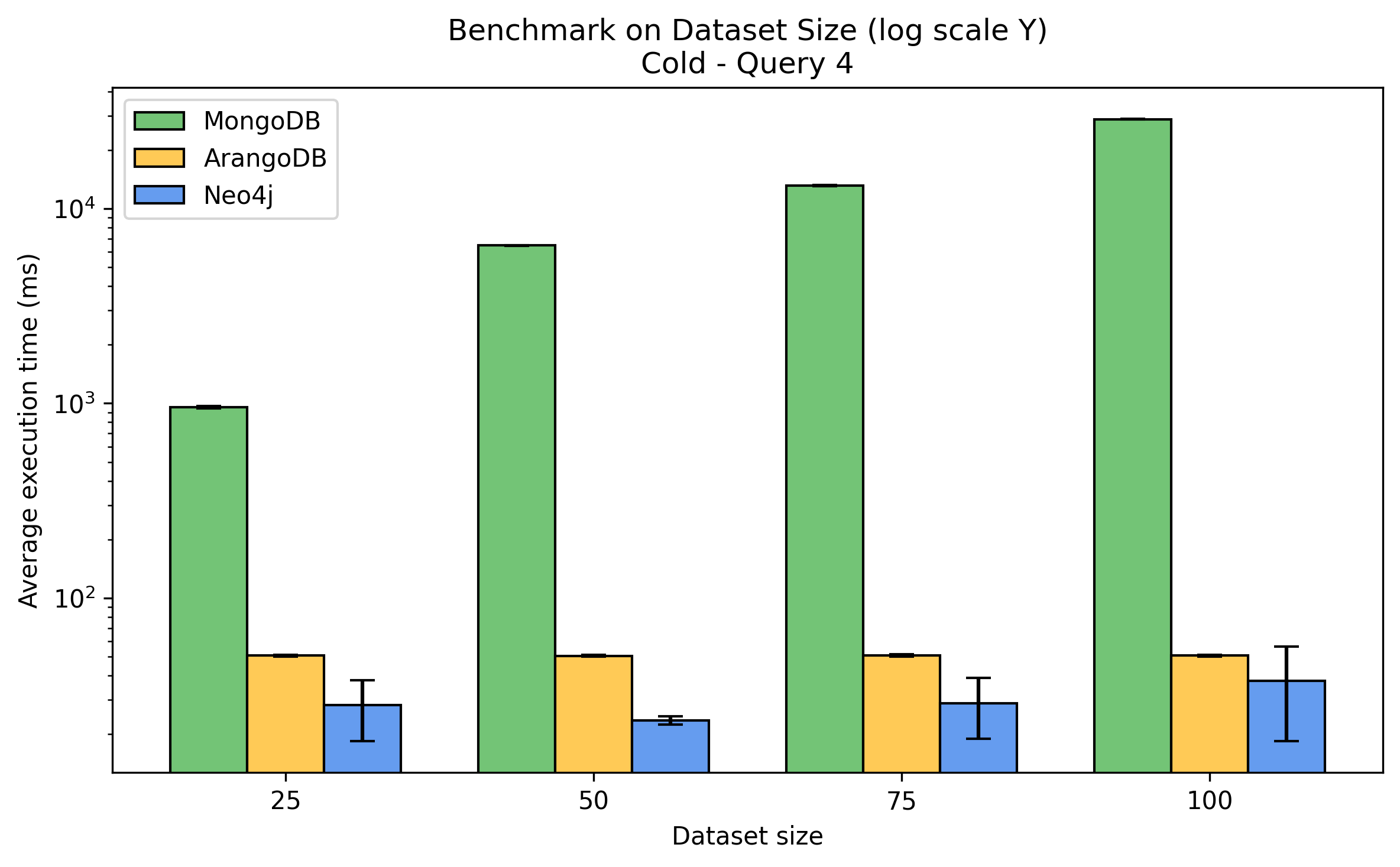}\\
            \footnotesize Cold
        \end{minipage}
        \hfill
        \begin{minipage}{0.48\columnwidth}
            \centering
            \includegraphics[width=\linewidth]{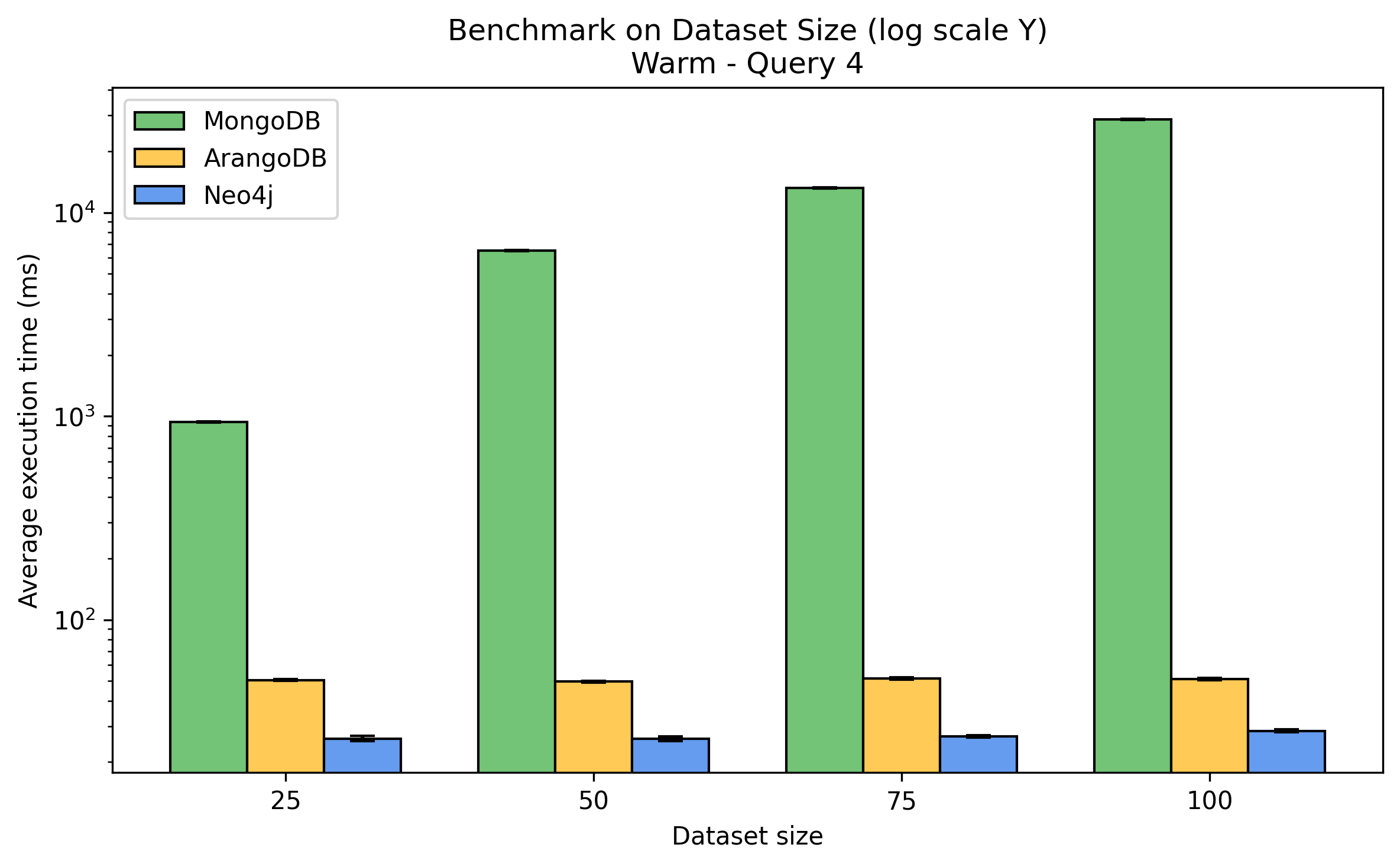}\\
            \footnotesize Warm
        \end{minipage}
    \end{minipage}
}
\caption{Query 4 Results in Distributed Cluster.}
\end{figure}

Q4 results represent the most pronounced manifestation of logical misalignment in this study. Neo4j sustains sub-20~ms latency across all dataset profiles with negligible variance, ArangoDB remains stable around 50~ms, while MongoDB grows from $10^3$~ms at 25\% to over $3 \times 10^4$~ms at full scale, a gap of more than three orders of magnitude. The LIMIT applied to starting nodes and results isolates structural pattern matching cost from serialization overhead, ensuring the observed gap is directly attributable to the data model. Neo4j resolves each hop via direct pointer dereference, keeping per-step cost independent of graph size, whereas MongoDB decomposes the 4-hop chain into nested \$lookup stages whose cost scales super-linearly with both traversal depth and dataset volume.

%% file: tex/discussion.tex
\section{Discussion}

The experimental results show that increasing the available computational resources is not a sufficient condition for improving performance in Big Data systems. When the adopted logical model is misaligned with the structure of the queries, adding nodes to the cluster yields performance equivalent to or worse than a single-machine configuration.

This phenomenon, systematically observable across all four query classes, highlights the decisive role of logical scalability over purely infrastructural scalability. The progressive nature of this effect is directly traceable in the results. In Q1, where no relational traversal is required, the document-oriented model operates within its design envelope and delivers the lowest execution times, confirming that logical alignment is beneficial in both directions, in particular a well-matched model scales efficiently, while a mismatched one degrades regardless of available resources. As query complexity increases through Q2 and Q3, the cost of simulating graph relationships through MongoDB's aggregation pipeline grows non-linearly, accumulating an overhead that no additional hardware can absorb, since the bottleneck is structural rather than computational. Q4 makes this conclusion unambiguous: Neo4j sustains sub-20~ms latency across all dataset scales, while MongoDB exceeds $3 \times 10^4$~ms at full scale, a gap of more than three orders of magnitude that persists independently of the resources allocated. This evidence supports a broader principle: in architectures where the logical model is misaligned with the workload, distributed scaling introduces additional coordination and communication overhead on top of an already inefficient retrieval strategy, compounding the performance penalty rather than mitigating it. 

The computational advantages expected from horizontal scaling are systematically neutralized by the structural bottlenecks of the data model, which becomes the binding constraint for the entire architecture. Logical scalability is therefore not a property that emerges automatically from infrastructural investment, it must be deliberately engineered through the selection of a data model whose internal structure reflects the inherent topology of the data and the retrieval patterns of the application.

%% file: tex/reproducibility.tex
\section*{Reproducibility Statement}

To ensure scientific transparency and allow for the full verification of the experimental results presented in this study, the entire software ecosystem developed for this research has been made publicly available at https://github.com/rosarionapoli99/big-data-logical-scalability